\documentclass[conference]{IEEEtran}
\IEEEoverridecommandlockouts
\usepackage{cite}
\usepackage{amsmath,amssymb,amsfonts}
\usepackage{algorithmic}
\usepackage{algorithm}
\usepackage{graphicx}
\usepackage{textcomp}
\usepackage{xcolor}
\usepackage{threeparttable}
\usepackage{pifont}
\usepackage{booktabs}
\usepackage{url}

\def\BibTeX{{\rm B\kern-.05em{\sc i\kern-.025em b}\kern-.08em
    T\kern-.1667em\lower.7ex\hbox{E}\kern-.125emX}}
\begin{document}

\title{Optimizing and Exploring System Performance in Compact
Processing-in-Memory-based Chips

\vspace{-12pt}
}

\author{Peilin Chen, Xiaoxuan Yang\\
Department of Electrical and Computer Engineering, University of Virginia, Charlottesville, VA, USA\\
\{peilin, xiaoxuan\}@virginia.edu
\vspace{-18pt}
}
\maketitle

\begin{abstract}

Processing-in-memory~(PIM) is a promising computing paradigm to tackle the ``memory wall" challenge. However, PIM system-level benefits over traditional von Neumann architecture can be reduced when the memory array cannot fully store all the neural network~(NN) weights. The NN size is increasing while the PIM design size cannot scale up accordingly due to area constraints. Therefore, this work targets the system performance optimization and exploration for compact PIM designs. We first analyze the impact of data movement on compact designs. Then, we propose a novel pipeline method that maximizes the reuse of NN weights to improve the throughput and energy efficiency of inference in compact chips. To further boost throughput, we introduce a scheduling algorithm to mitigate the pipeline bubble problem. Moreover, we investigate the trade-off between the network size and system performance for a compact PIM chip. Experimental results show that the proposed algorithm achieves $2.35\times$ and $0.5\%$ improvement in throughput and energy efficiency, respectively. Compared to the area-unlimited design, our compact chip achieves approximately $56.5\%$ of the throughput and $58.6\%$ of the energy efficiency while using only one-third of the chip area, along with $1.3\times$ improvement in area efficiency. Our compact design also outperforms the modern GPU with $4.56\times$ higher throughput and $157\times$ better energy efficiency. Besides, our compact design uses less than $20\%$ of the system energy for data movement as batch size scales up. 

\end{abstract}

\section{Introduction}

Convolutional neural networks~(CNNs) have gained great success in various applications such as image recognition and autonomous driving~\cite{ref23, ref24, ref25}. To improve training and inference accuracy, the complexity and size of CNNs keep increasing. The large-scale parameters of CNNs generate substantial data movement and computation load. Even though prior studies focus on optimizing the computation part, data movement contributes to a significant portion~($62.7\%$) of the total energy consumption~\cite{boroumand2018google}. To tackle this challenge, researchers propose processing-in-memory~(PIM) architecture~\cite{ref1, ref37} to reduce the data movement overhead by storing the data within the computing unit. However, PIM-based chips still face significant challenges when applied to NN inference due to the trade-off between NN model size and available chip area.

\begin{figure}[tb]
    \centering
    \includegraphics[width=0.85\linewidth]{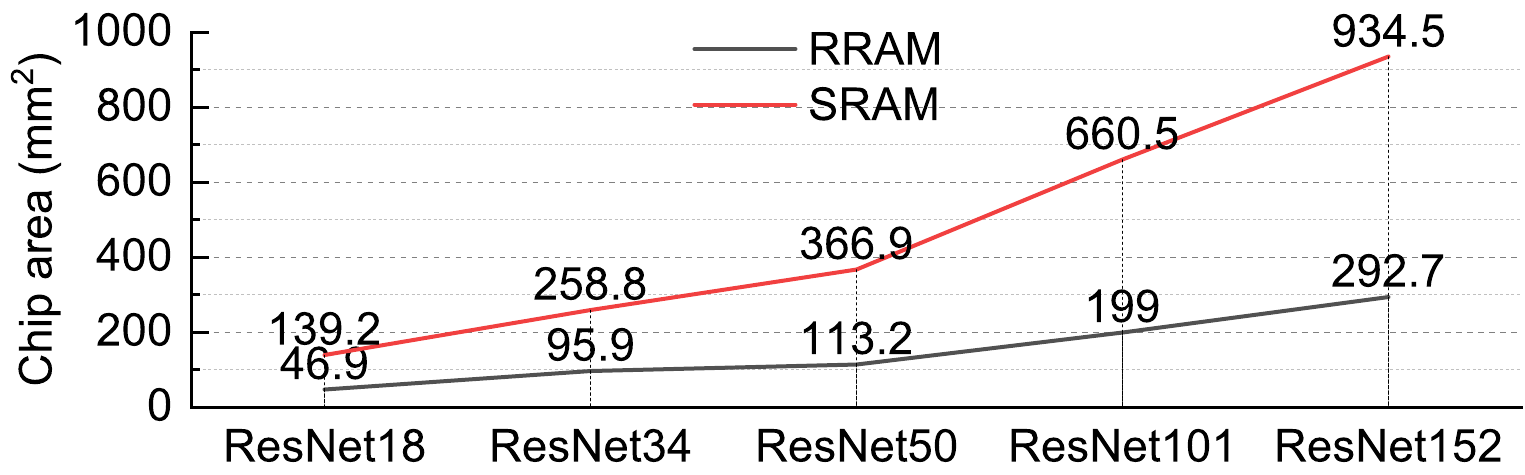}
    \caption{Chip area required by ResNet NNs to store all weights on SRAM and RRAM array under 32nm process. Such designs are referred to as area-unlimited designs.\vspace{-12pt}}
    \label{fig1}
    \vspace{-6pt}
\end{figure}

Firstly, prior works~\cite{chih202116, dong202015, jiang202240nm, tu2022redcim} overlook the impact of off-chip data movement in their system evaluation of compact~(area-limited) PIM designs. As shown in Fig.~\ref{fig1}, to put all weights of ResNet152 with $58$ million parameters onto SRAM and resistive random access memory~(RRAM) designs, the chip area is scaled up to $934.5mm^2$ and $292.7mm^2$. It is impractical to design such large PIM-based chips given the practical area and cost budget~\cite{ref11}. In the PIM processor design~\cite{tu2022redcim}, we find that the small-scale design, with only 12KB PIM capacity, cannot store all weights of NNs. However, the reporting results only consider the energy consumption of all on-chip components, excluding off-chip memory. Integrating off-chip memory for system evaluation is important for understanding the impact of data movement on compact chips.

Secondly, the compact PIM-based chips suffer from poor inference performance compared to impractical area-unlimited designs. Due to the practical assumption of area constraint, compact chips can only load and store a fixed portion of weights each time. It is unavoidable for compact chips to repeatedly erase the current weights and reload other weights. The limited resources hinder compact chips from fully achieving the optimal chip performance. The original pipeline designed for better throughput in area-unlimited chips cannot work effectively on compact chips. Developing techniques specifically tailored to compact PIM-based chips is essential to bridging this performance gap.

Thirdly, for a fixed-scale PIM-based chip, there is a trade-off between the size of deployed CNNs and the performance of leveraging PIM designs. Unavoidable data movement might offset the advantage of compact PIM designs. However, due to PIM advantage in the high parallelism and low energy consumption, there is an upper limit to the size of CNNs that can maintain the performance of PIM-based chips with minimal impact. Therefore, we aim to investigate the trade-off between the deployed NN size and system performance.

Based on these challenges, our work aims to optimize and explore system performance in compact PIM-based chips. The main contributions of our work are as follows.

\begin{enumerate}
    \item We first analyze the impact of data movement on the performance of compact PIM-based chips. Then, we propose a novel pipeline method to improve the throughput and energy efficiency of inference in compact chips. 
    \item To further boost throughput, we design a scheduling algorithm to speed up the bottleneck layers dynamically and mitigate the pipeline bubble problem.
    \item We explore the maximum size of CNNs that can be deployed onto a given compact PIM-based chip while maintaining its advantage. 
\end{enumerate}

\section{Methodology}
\label{section3}

\subsection{Overall workflow}

\begin{figure}[tb]
    \centering
    \includegraphics[width=\linewidth]{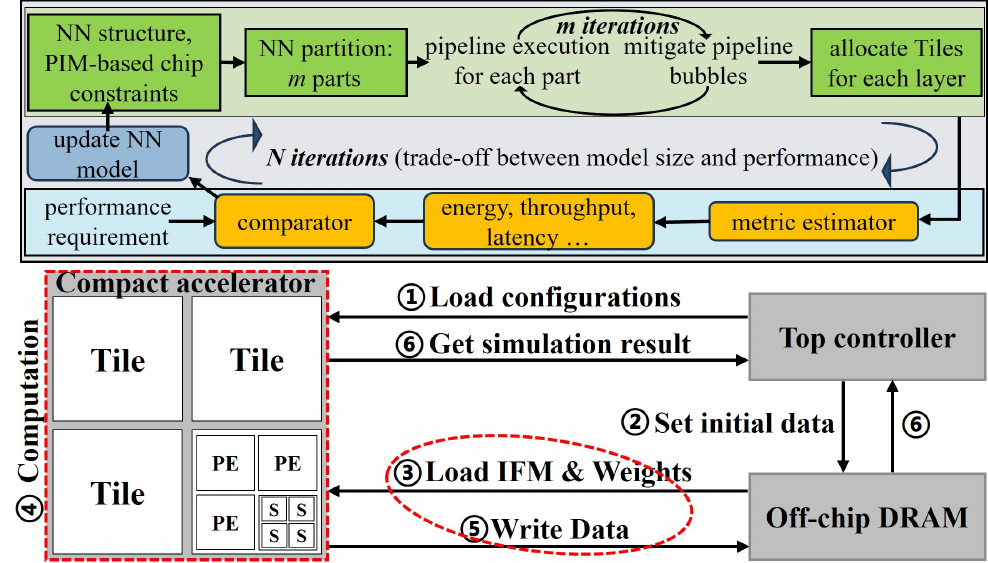}
    \caption{Overall workflow of our design. PE: Processing Engine. S: Subarray.}
    \label{fig2}
\end{figure}

We present the overall architecture and workflow of our design in Fig.~\ref{fig2}. We build a compact PIM-based accelerator~(denoted in the red dotted box) for evaluation. To ensure fair comparison and accurate benchmarking of system performance, we consider the energy consumption and latency of data movement using different types and specifications of off-chip DRAM, such as LPDDR3~\cite{ref26}, LPDDR4~\cite{ref27}, and LPDDR5~\cite{ref28}. We precisely record the data movement in steps \ding{174} and \ding{176} in the following format: \textit{transaction time}, \textit{transaction type}~(write/read), \textit{logical memory address}~(32-bit). Moreover, to improve system throughput and reduce the impact of data movement, we propose a novel pipeline method tailored for compact chips and design a scheduling algorithm to mitigate the pipeline bubbles. The search iteration in the upper part covers the NN partition, our proposed designs, resource allocation, metrics evaluation, etc.

\subsection{The Impact of Data Movement on Compact PIM Designs}

\begin{figure}[tb]
    \centering
    \includegraphics[width=0.7\linewidth]{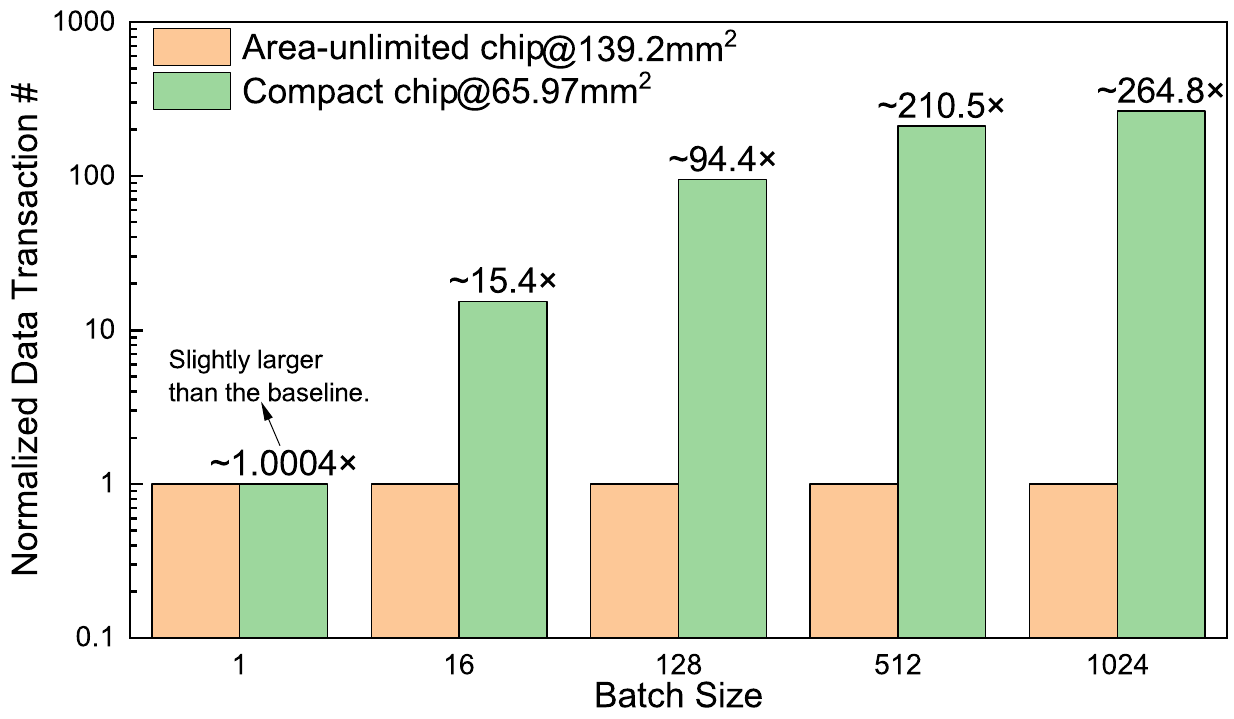}
    \caption{Normalized data transaction number for different batch sizes between PIM designs and LPDDR5.}
    \label{fig3}
\end{figure}

Data movement can be a severe challenge in compact PIM-based designs. To better understand this issue, we evaluate the data movement of ResNet18 across different batch sizes in the compact chip and the area-unlimited chip that can store all weights of ResNet18, as shown in Fig.~\ref{fig3}. When batch size scales up to $1024$, the data movement number for the compact chip is $264.8\times$ larger than the area-unlimited design. The compact chip repeatedly reloads both intermediate data and NN weights, while all weights remain fixed in the area-unlimited chip. This is the major factor that significantly offsets the PIM advantage in the compact design. To tackle this challenge, we propose a novel pipeline method tailored to compact chip, design a scheduling algorithm to mitigate the pipeline bubble, and explore the maximum NN size that can deployed in compact chips while maintaining its advantage.

\subsection{Pipeline Method for Compact PIM Designs}
\label{section3_2}

\begin{figure}[tb]
    \centering
    \includegraphics[width=\linewidth]{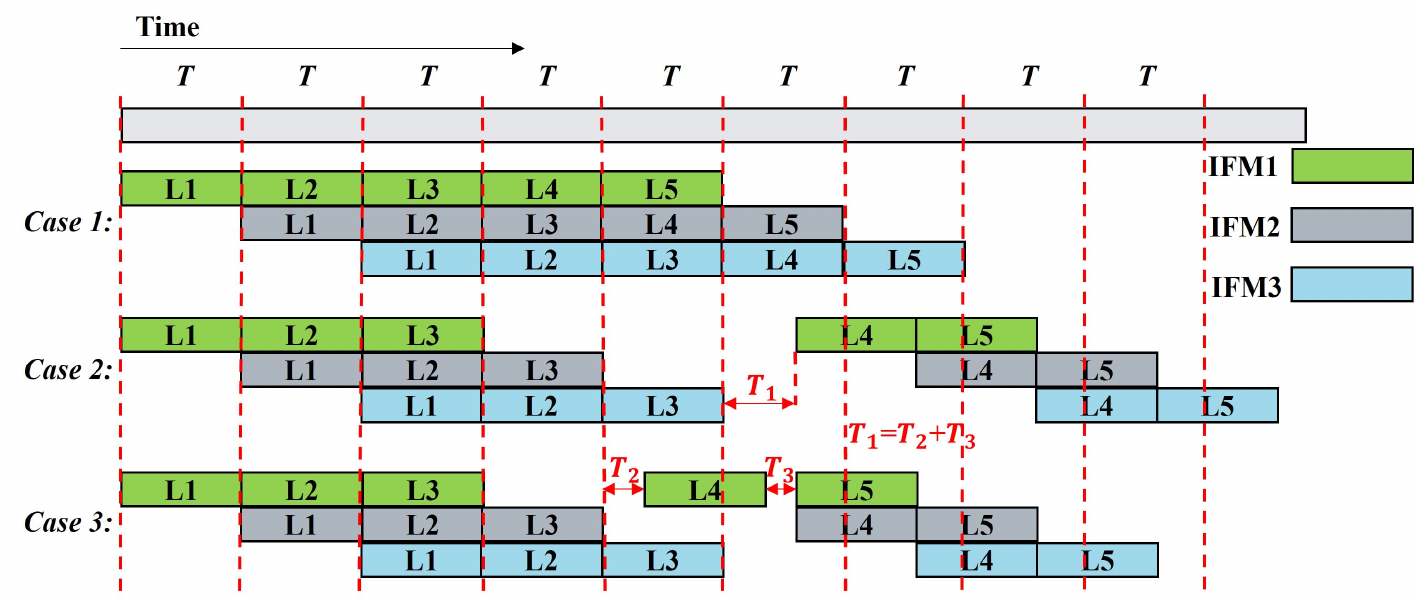}
    \caption{Pipeline method in area-unlimited designs~(case 1) vs. our pipeline method in compact PIM-based chips~(case 2 and case 3). In this example, we assume that the NN has five CONV/FC layers~(L1$\sim$L5). }
    \label{fig4}
\end{figure}

Compact PIM-based chips cannot store all the weights of large CNNs. Therefore, for the compact chip, we must design specific principles to partition the NNs. Intuitively, partitioning the NNs by layers can preserve the NN structure while reducing the storage requirement. To reduce the memory access, we follow partition criteria that allow the chip to map as many layers as possible for each loading process. If the available chip area cannot store one single layer's weights, we would further partition the layer's weight along the input and output channels, which also aligns with work~\cite{ref16}. In summary, our method partitions by layer based on the available storage size and further partitions by channels if necessary.

To boost throughput and energy efficiency in compact PIM-based chips, we propose a novel pipeline method in our design. Fig.~\ref{fig4} shows the difference between the pipeline method in the area-unlimited~(where the chip can store all the NN weights) and our proposed pipeline method in compact PIM designs. In Fig.~\ref{fig4}, we assume that different layers consume the same time \textit{T}~(\textit{T} is longer than one cycle) during inference. For case 1 in Fig.~\ref{fig4}, we can formulate the time consumption as follows:
\begin{equation*}
    t(n)_{case1}=\textit{L}\times\textit{T}+(n-1)\textit{T}=(n+\textit{L}-1)\textit{T}
\end{equation*}
\begin{equation*}
    t(perIFM)_{case1}= \frac{t(n)_{case1}}{n}= \frac{(n+\textit{L}-1)\textit{T}}{n} \approx \textit{T}, 
\end{equation*}
where \textit{L} and \textit{n} denote the layer number and the number of input feature map~(IFM) or batch size, and \textit{t(n)} and \textit{t(perIFM)} represent the inference time of \textit{n} IFMs and per IFM, respectively. When there are continuous inputs~(or the batch size is large), the chip consumes an average latency of \textit{T} to complete the inference of each IFM. Compact PIM designs cannot achieve the same optimization effectiveness as in case 1 due to the limited computation resources. However, we propose a novel pipeline method~(case 2 and case 3 in Fig.~\ref{fig4}) to improve throughput and energy efficiency in compact PIM designs, reducing the performance gap between the area-unlimited designs and compact chips. The \textit{t(perIFM)} for case 2 can be calculated as follows:
\begin{equation*}
    t(perIFM)_{case2} = \frac{t(n)_{case2}}{n}=\frac{(2n+L-2)\textit{T}+T_1}{n},
\end{equation*}
where \textit{$T_1$} denotes the latency for loading the intermediate data and the weights of L4 and L5. The NN in case 2 is divided into two parts. The compact chip performs inference on IFMs in a pipeline manner within each part. By setting a suitable batch size \textit{n} that considers the latency to get the inference result, we can make each part consume the same time as $t(perIFM)_{case1}$. Therefore, $t(perIFM)_{case2}$ is equal to $2\textit{T}$. Note that \textit{n} in case 2 cannot be infinite because the inference would always remain on the first part and never proceed to the second part. Moreover, we show a special pipeline situation in our design~(case 3). Part 2 can ideally be processed \textit{T} earlier if L4 in part 2 requires no more memory capacity than L1 and L2 in part 1, and if the memory requirement for part 2 is larger than that for L1 and L2 in part 1. Different from case 2, case 3 first loads the intermediate data and the weights of L4 into the chip~(latency: \textit{$T_2$}). After both blue L3 and green L4 finish computation, we begin to load the weights of L5~(latency: $T_3$). In summary, $t(perIFM)_{case3}$ is equal to $\frac{(2n+L-1)\textit{T}+T_2+T_3}{n}$.

\subsection{Dynamic Duplication Method}
\label{section3_3}

\begin{algorithm}[tb]
    \caption{Dynamic duplication method (DDM)}
    \label{algorithm1}
    \renewcommand{\algorithmicrequire}{\textbf{Input:}}
    \renewcommand{\algorithmicensure}{\textbf{Output:}}
    
    \begin{algorithmic}[1]
        \REQUIRE \textit{$N_{tile}[i]$}: Number of \textit{Tile} required by each layer; \textit{E}: Extra number of \textit{Tile} after mapping a portion of the NN weights onto the chip; \textit{MAX[i]}: Maximum duplication number for each layer; \textit{N}: Total number of \textit{Tile}.
        \ENSURE \textit{dupNum[i]}: Duplication number of each layer.

        \STATE Divide NN into \textit{m} parts and partition the $i^{th}$ layer if $N_{tile}[i]>\textit{N}$; Initialize inference time predictor (ITP);
        \FOR{each $i \in m-1$}
            \STATE Initialize the \textit{Flag} and record the minimum \textit{Tile} numbers (\textit{$min_{tile}$}) among all layers in this part;
            \WHILE{$\textit{E} \geq \textit{$min_{tile}$}$}
                \STATE Update ITP and select bottleneck layer \textit{l};
                \IF {$E \geq N_{tile}[l]$}
                    \STATE \textit{Flag}=1; \textit{E}=\textit{E}-\textit{$N_{tile}[l]$}; \textit{dupNum[l]}=\textit{dupNum[l]}+1;
                    \IF{FC layer} 
                        \STATE \textit{dupNum[l]}=1; \textit{Flag}=0;
                    \ELSIF{$\textit{dupNum[l]} > \textit{MAX[l]}$}
                        \STATE \textit{dupNum[l]}=\textit{dupNum[l]}-1; \textit{Flag}=0;
                    \ENDIF
                \ELSE
                    \STATE \textit{Flag}=0;
                \ENDIF
            \ENDWHILE
        \ENDFOR
        
    \end{algorithmic}
\end{algorithm}

\begin{figure}[tb]
    \centering
    \includegraphics[width=0.81\linewidth]{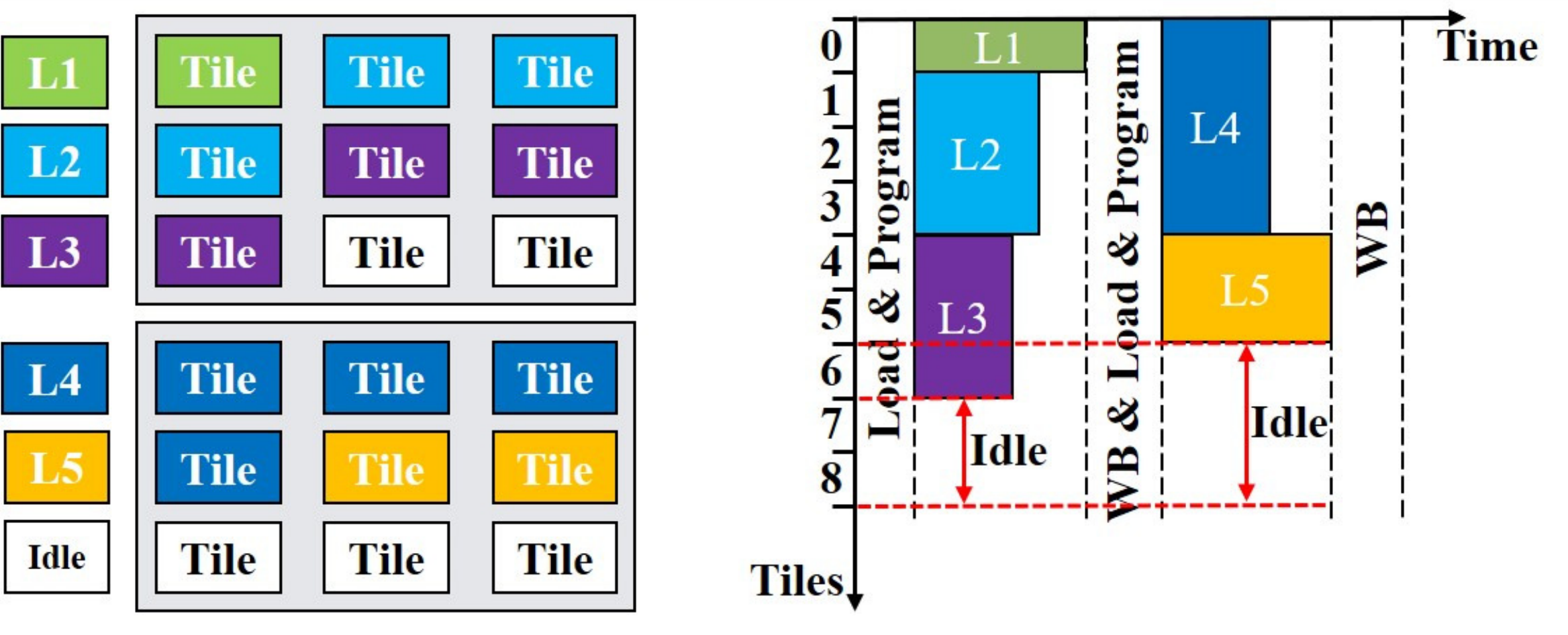}
    \caption{Left: Five NN layers are divided into two parts to be mapped onto the accelerator. Right: Execution process of the NN. WB denotes ``Write Back''.}
    \label{fig5}
\end{figure}

In Section~\ref{section3_2}, we assume that each NN layer consumes the same inference time, which is ideal but impractical. The layer with the longest latency will appear as the bottleneck in the pipeline. We need to mitigate the pipeline bubble in order to achieve higher throughput. Therefore, we propose an algorithm to dynamically speed up the bottleneck layers. In our design, we assume that one \textit{Tile} is the minimum computational unit, and mapping more than one layer onto the same \textit{Tile} is not allowed. Due to this assumption, it is feasible to duplicate layers at the \textit{Subarray}, \textit{PE}, and even \textit{Tile} levels. As shown in Fig.~\ref{fig5}, there are some idle \textit{Tiles} after mapping selected layers onto the chip. Thus, we can utilize unused \textit{Tiles} to accelerate the bottleneck layers. Algorithm~\ref{algorithm1} shows the details of the dynamic duplication method (DDM) with our pipeline strategy. We begin by partitioning NN according to the criteria mentioned in Section~\ref{section3_2}. Based on the Roofline Model~\cite{ref20}, we observe that the inference time of each layer in PIM designs is proportional to the size of the output feature map~(OFM)~($O\times O$). Then, we initialize the inference time predictor based on this observation. Following that, we build a \textit{for} loop with \textit{m} iterations~(NN is divided into \textit{m} parts) to search for the optimal duplication strategy for each partition part. Note that the maximum duplication value for each layer~(\textit{MAX[i]}) is also related to $O \times O$. For example, if \textit{O}=8, we can duplicate this layer up to $64$ times, meaning this layer can be computed within one cycle~\cite{ref9}. The \textit{Flag} determines whether to skip a layer that cannot be duplicated. The \textit{While} loop continues if \textit{E}~(extra number of \textit{Tiles} after mapping) is greater than the minimum number of \textit{Tiles} required among all layers. Finally, this algorithm returns \textit{dupNum}, which records the duplication number for each layer.

\section{Experimental results and analysis}
\label{section4}

\subsection{Experiment Setup}

We build our evaluation framework based on the NeuroSim for PIM-based designs~\cite{ref2} and the DRAMPower~\cite{ref5} for data movement. We design the top controller to support our NN partition strategy, layer duplication, novel pipeline method, and dynamic duplication for bottleneck layers. We select several CNNs to deploy on the compact chip for evaluation, including ResNet-18, ResNet-34, ResNet-50, ResNet-101, and ResNet-152~\cite{ref33} for CIFAR-100 dataset~\cite{ref30}. The weights and activations of NN are quantized to 8-bit~\cite{ref31}. We use NeuroSim to benchmark the inference performance of area-unlimited designs. Note that NeuroSim does not explicitly include the data movement of the IFM, weights, and OFM. For fair comparison, both the area-unlimited chip and our compact design use 8Gb 4266MHz 128-bit LPDDR5 as DRAM~\cite{ref28}.

\subsection{Throughput and Energy Efficiency Improvement}

\label{section4_2}

\begin{figure}[tb]
    \centering
    \includegraphics[width=0.86\linewidth]{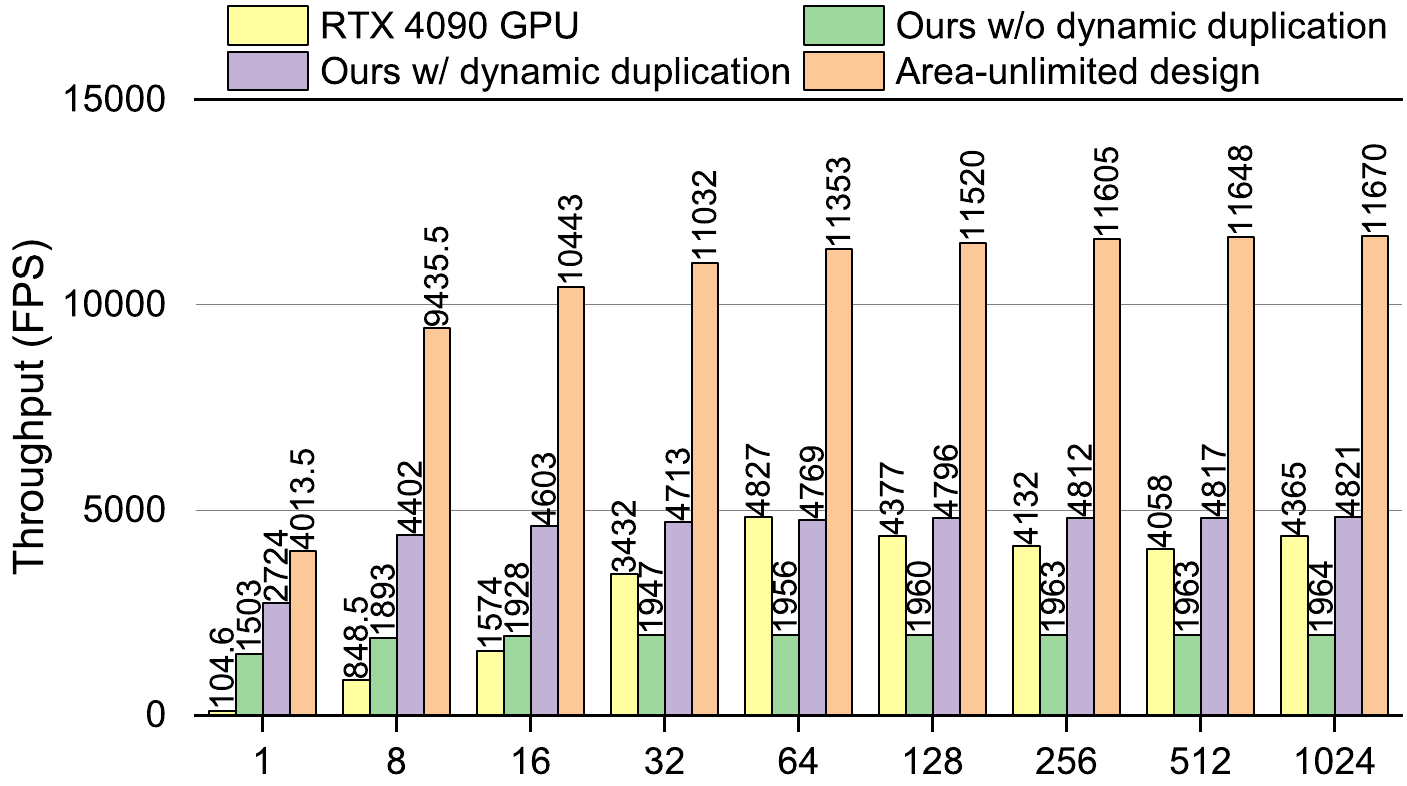}
    \includegraphics[width=0.86\linewidth]{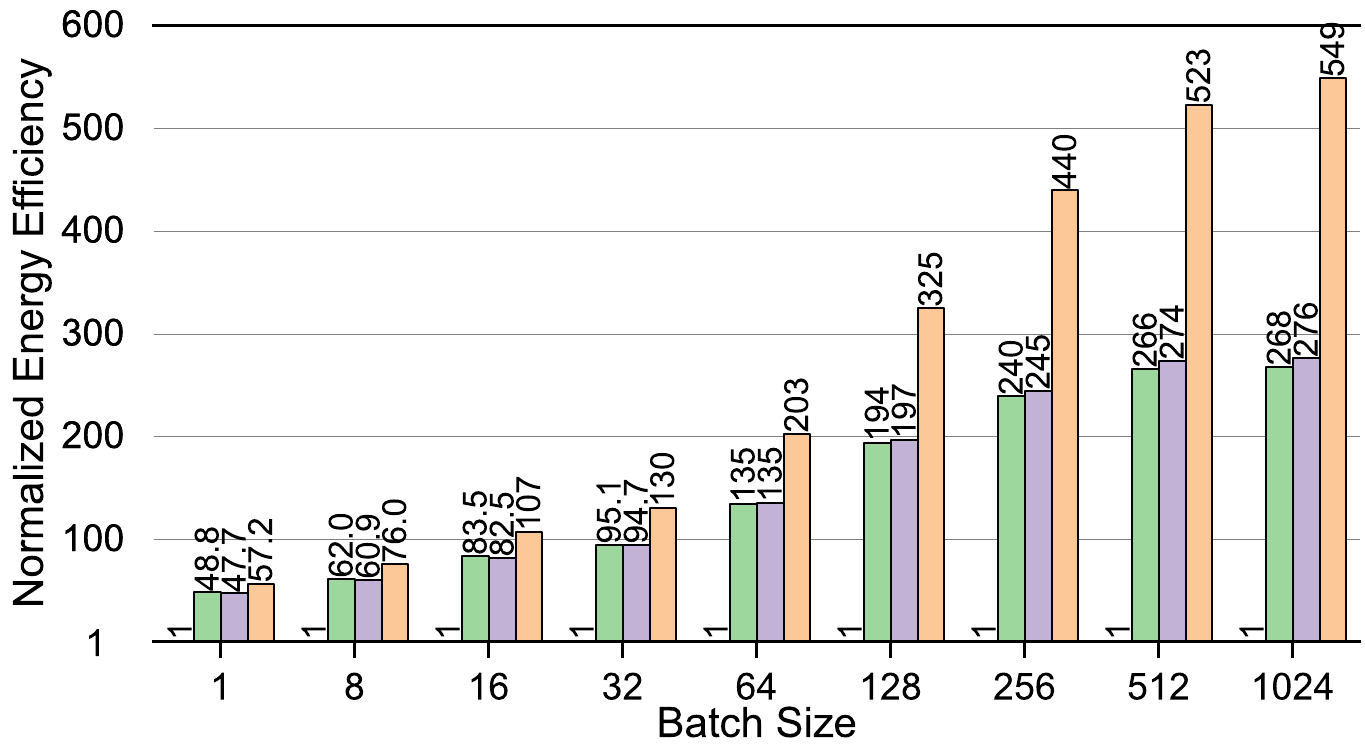}
    \caption{Evaluating throughput and energy efficiency under different batch sizes: RTX 4090 GPU, our design without and with dynamic duplication method, and area-unlimited chip.}
    \label{fig6}
    \vspace{-10pt}
\end{figure}

We present the throughput and energy efficiency results for ResNet-34 inference using RTX 4090 GPU, our compact design with an area of $41.5mm^2$, and the area-unlimited chip with an area of $123.8mm^2$ in Fig.~\ref{fig6}. Firstly, compared to the RTX 4090 GPU, our design with DDM achieves $4.56\times$ improvement in throughput and $157\times$ improvement in energy efficiency. Secondly, our design with DDM achieves approximately $56.5\%$ of the throughput and $58.6\%$ of the energy efficiency while using only one-third of the chip area compared to the area-unlimited chip. Thirdly, the DDM achieves $2.35\times$ and $0.5\%$ improvement in throughput and energy efficiency compared to the design without DDM. The reason is twofold. On the one hand, we accelerate the bottleneck layers to mitigate the pipeline bubble, thus enabling $2.35\times$ throughput improvement. On the other hand, even though the DDM introduces additional memory access and computation, the benefit of saving leakage energy~(due to fewer idle cycles) is more significant. Therefore, energy efficiency gains slightly. Finally, our design with DDM achieves an average of $16.2GOPS/mm^2$ across all batch sizes, which surpasses the area-unlimited design's $12.5GOPS/mm^2$. This highlights our compact design in area efficiency.

\subsection{Data Movement and Energy Breakdown}

\label{section4_3}

\begin{figure}[tb]
    \centering
    \includegraphics[width=\linewidth]{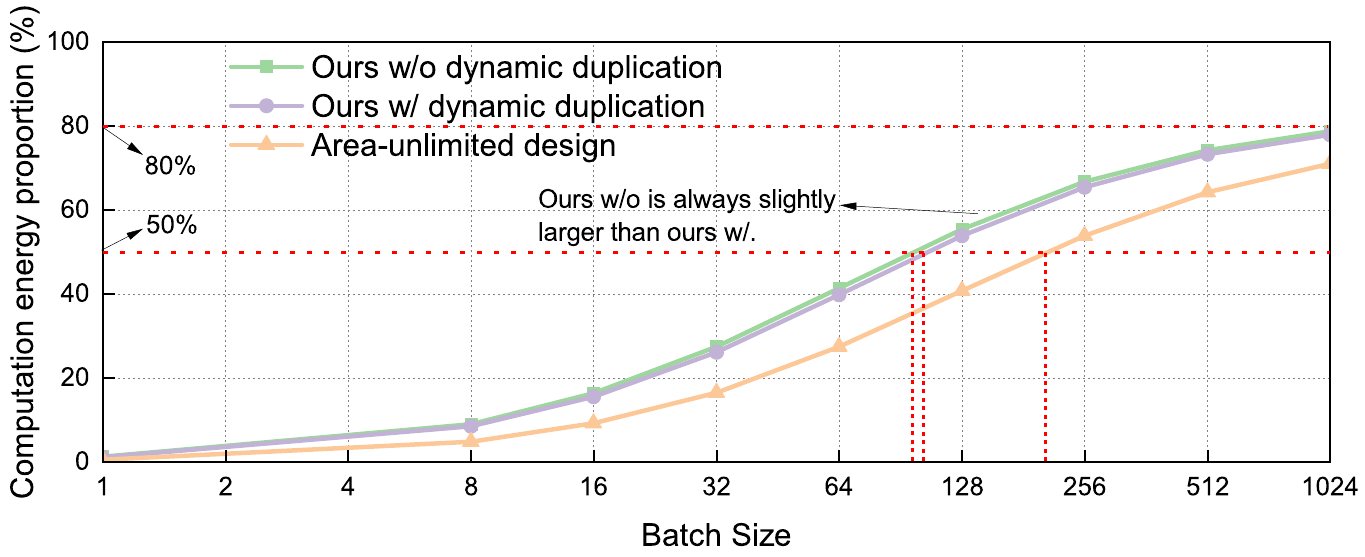}
    \caption{Computation energy proportion over different batch sizes. Total energy consists of computation energy and off-chip DRAM energy.}
    \label{fig7}
    \vspace{-10pt}
\end{figure}

We further show the proportion of computation energy in the total energy consumption for our design and the area-unlimited chip in Fig.~\ref{fig7}. We separate the total energy consumption into two parts: 1) computation energy~(which refers to the computation energy of all on-chip components) and 2) off-chip DRAM energy. As the batch size increases, the computation energy occupies more than $50\%$, and even up to $80\%$, of the entire system energy. Thus, off-chip DRAM energy does not have a significant impact on system energy efficiency in our design if a suitable batch size is selected for NN inference. 

\subsection{Maximum NN Size Deployment in Compact PIM Design}

\begin{figure}[tb]
    \centering
    \includegraphics[width=\linewidth]{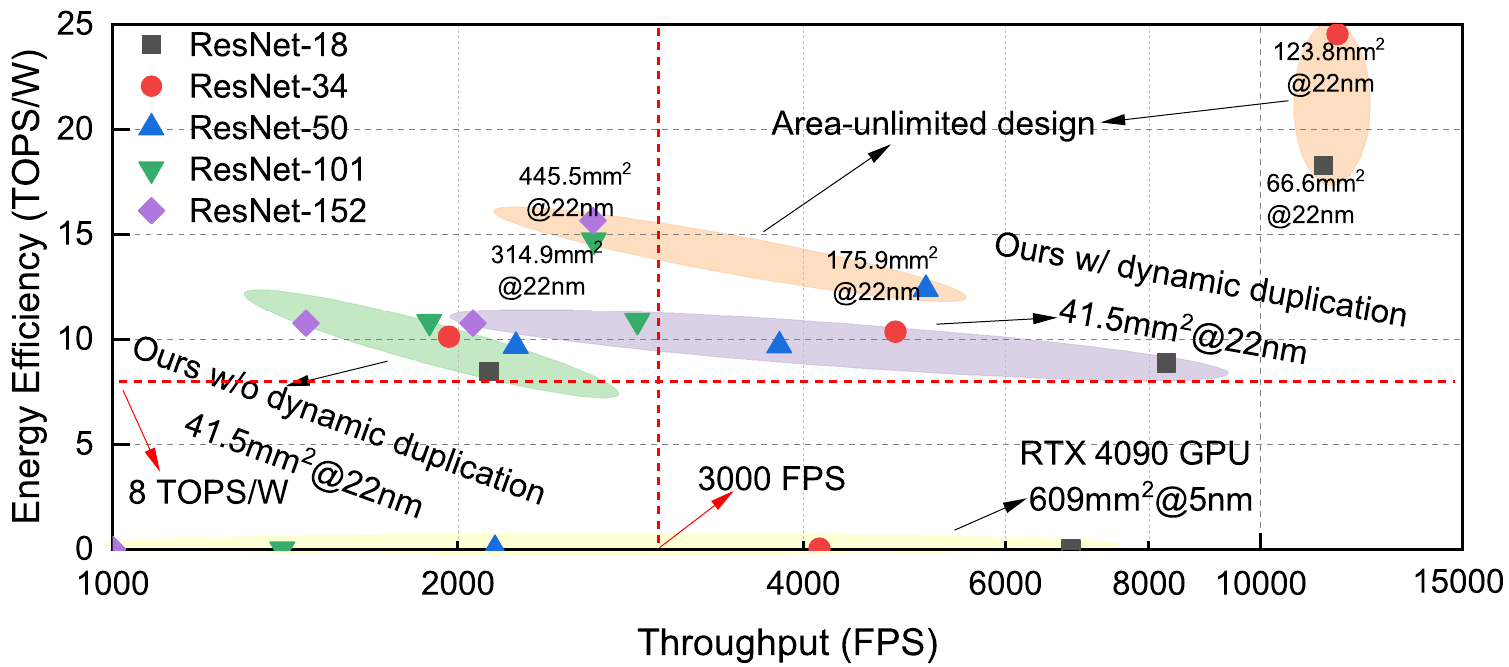}
    \caption{Exploration of maximum NN size deployed in our compact chip.}
    \label{fig8}
    \vspace{-10pt}
\end{figure}

The maximum NN size exploration results are presented in Fig~\ref{fig8}. We can observe that the area-unlimited design has obvious advantages over our compact chip due to its sufficient area. However, the DDM narrows the performance gap between our pipeline design and the area-unlimited chip. We focus on our design with dynamic duplication to analyze the exploration of maximum NN size deployment~(highlighted by the purple oval). The inference throughput decreases rapidly as the NN grows larger. At the same time, our design has slight fluctuations in energy efficiency but maintains over 8 TOPS/W for all ResNet NNs. For a specific performance requirement~(such as $\text{energy efficiency} \geq \text{8 TOPS/W}$ and $\text{throughput} \geq \text{3000 FPS}$), it is recommended to deploy ResNet NNs that are smaller than ResNet-101 in our compact chip. Note that this is a rough estimation due to the network's coverage in the current series of experiments. But the maximum size of the ResNet NN that can be deployed onto our compact chip ranges between ResNet-50~(23.7M) and ResNet-101~(42.6M).

\section{Conclusion}
\label{section5}

In this paper, we analyze the impact of data movement on the performance of compact designs and propose a novel pipeline method to improve throughput and energy efficiency. To address the pipeline bubble issue, we develop a scheduling algorithm to speed up the bottleneck layers dynamically. Additionally, we explore and provide insights into the maximum NN size that can be deployed in our compact PIM-based chip.

\bibliographystyle{ieeetr}
\bibliography{citations}

\begin{thebibliography}{10}

\bibitem{ref23}
A.~Krizhevsky, I.~Sutskever, and G.~E. Hinton, ``Imagenet classification with deep convolutional neural networks,'' {\em Advances in neural information processing systems}, vol.~25, 2012.

\bibitem{ref24}
V.~Sze, Y.-H. Chen, T.-J. Yang, and J.~S. Emer, ``Efficient processing of deep neural networks: A tutorial and survey,'' {\em Proceedings of the IEEE}, vol.~105, no.~12, pp.~2295--2329, 2017.

\bibitem{ref25}
L.~Deng, G.~Li, S.~Han, L.~Shi, and Y.~Xie, ``Model compression and hardware acceleration for neural networks: A comprehensive survey,'' {\em Proceedings of the IEEE}, vol.~108, no.~4, pp.~485--532, 2020.

\bibitem{boroumand2018google}
A.~Boroumand, S.~Ghose, Y.~Kim, R.~Ausavarungnirun, E.~Shiu, R.~Thakur, D.~Kim, A.~Kuusela, A.~Knies, P.~Ranganathan, {\em et~al.}, ``Google workloads for consumer devices: Mitigating data movement bottlenecks,'' in {\em Proceedings of the Twenty-Third International Conference on Architectural Support for Programming Languages and Operating Systems}, pp.~316--331, 2018.

\bibitem{ref1}
X.~Yang, B.~Taylor, A.~Wu, Y.~Chen, and L.~O. Chua, ``Research progress on memristor: From synapses to computing systems,'' {\em IEEE Transactions on Circuits and Systems I: Regular Papers}, vol.~69, no.~5, pp.~1845--1857, 2022.

\bibitem{ref37}
N.~Verma, H.~Jia, H.~Valavi, Y.~Tang, M.~Ozatay, L.-Y. Chen, B.~Zhang, and P.~Deaville, ``In-memory computing: Advances and prospects,'' {\em IEEE Solid-State Circuits Magazine}, vol.~11, no.~3, pp.~43--55, 2019.

\bibitem{chih202116}
Y.-D. Chih, P.-H. Lee, H.~Fujiwara, Y.-C. Shih, C.-F. Lee, R.~Naous, Y.-L. Chen, C.-P. Lo, C.-H. Lu, H.~Mori, {\em et~al.}, ``16.4 an 89tops/w and 16.3 tops/mm 2 all-digital sram-based full-precision compute-in memory macro in 22nm for machine-learning edge applications,'' in {\em 2021 IEEE International Solid-State Circuits Conference (ISSCC)}, vol.~64, pp.~252--254, IEEE, 2021.

\bibitem{dong202015}
Q.~Dong, M.~E. Sinangil, B.~Erbagci, D.~Sun, W.-S. Khwa, H.-J. Liao, Y.~Wang, and J.~Chang, ``15.3 a 351tops/w and 372.4 gops compute-in-memory sram macro in 7nm finfet cmos for machine-learning applications,'' in {\em 2020 IEEE International Solid-State Circuits Conference-(ISSCC)}, pp.~242--244, IEEE, 2020.

\bibitem{jiang202240nm}
H.~Jiang, W.~Li, S.~Huang, and S.~Yu, ``A 40nm analog-input adc-free compute-in-memory rram macro with pulse-width modulation between sub-arrays,'' in {\em 2022 IEEE Symposium on VLSI Technology and Circuits (VLSI Technology and Circuits)}, pp.~266--267, IEEE, 2022.

\bibitem{tu2022redcim}
F.~Tu, Y.~Wang, Z.~Wu, L.~Liang, Y.~Ding, B.~Kim, L.~Liu, S.~Wei, Y.~Xie, and S.~Yin, ``Redcim: Reconfigurable digital computing-in-memory processor with unified fp/int pipeline for cloud ai acceleration,'' {\em IEEE Journal of Solid-State Circuits}, vol.~58, no.~1, pp.~243--255, 2022.

\bibitem{ref11}
A.~Lu, X.~Peng, and S.~Yu, ``Compute-in-rram with limited on-chip resources,'' in {\em 2021 IEEE 3rd International Conference on Artificial Intelligence Circuits and Systems (AICAS)}, pp.~1--4, IEEE, 2021.

\bibitem{ref26}
Micron, ``Mobile lpddr3 sdram.'' \url{https://www.micron.com/-/media/client/global/documents/products/data-sheet/dram/mobile-dram/low-power-dram/lpddr3/178b_8-16gb_2c0f_mobile_lpddr3.pdf}, 2014.

\bibitem{ref27}
Micron, ``Lpddr4/lpddr4x sdram.'' \url{https://www.micron.com/-/media/client/global/documents/products/data-sheet/dram/mobile-dram/low-power-dram/lpddr4/z19m_embedded_lpddr4_lpddr4x.pdf}, 2022.

\bibitem{ref28}
{JEDEC Solid State Technology Association}, ``Low power double data rate (lpddr) 5/5x.'' \url{https://www.jedec.org/sites/default/files/docs/JESD209-5C.pdf}, 2023.

\bibitem{ref16}
K.~He, I.~Chakraborty, C.~Wang, and K.~Roy, ``Design space and memory technology co-exploration for in-memory computing based machine learning accelerators,'' in {\em 2022 IEEE/ACM International Conference On Computer Aided Design (ICCAD)}, pp.~1--9, 2022.

\bibitem{ref20}
S.~Williams, ``Roofline: An insightful visual performance model for floating-point programs and multicore,'' {\em ACM Communications}, p.~16, 2009.

\bibitem{ref9}
L.~Song, X.~Qian, H.~Li, and Y.~Chen, ``Pipelayer: A pipelined reram-based accelerator for deep learning,'' in {\em 2017 IEEE international symposium on high performance computer architecture (HPCA)}, pp.~541--552, IEEE, 2017.

\bibitem{ref2}
P.-Y. Chen, X.~Peng, and S.~Yu, ``Neurosim: A circuit-level macro model for benchmarking neuro-inspired architectures in online learning,'' {\em IEEE Transactions on Computer-Aided Design of Integrated Circuits and Systems}, vol.~37, no.~12, pp.~3067--3080, 2018.

\bibitem{ref5}
K.~Chandrasekar, C.~Weis, Y.~Li, B.~Akesson, N.~Wehn, and K.~Goossens, ``Drampower: Open-source dram power and energy estimation tool.'' \url{http://www.drampower.info}.

\bibitem{ref33}
K.~He, X.~Zhang, S.~Ren, and J.~Sun, ``Deep residual learning for image recognition,'' in {\em Proceedings of the IEEE conference on computer vision and pattern recognition}, pp.~770--778, 2016.

\bibitem{ref30}
A.~Krizhevsky, G.~Hinton, {\em et~al.}, ``Learning multiple layers of features from tiny images,'' 2009.

\bibitem{ref31}
S.~Wu, G.~Li, F.~Chen, and L.~Shi, ``Training and inference with integers in deep neural networks,'' in {\em International Conference on Learning Representations}, 2018.

\end{thebibliography}

\end{document}